# Sporadic Aurora near Geomagnetic Equator: In the Philippines, on 27 October 1856

Hisashi Hayakawa[1*, 2], José. M. Vaquero[3], and Yusuke Ebihara[4, 5]

[1]Graduate School of Letters, Osaka University, Toyonaka, 5600043, Japan (JSPS Research Fellow)
[2] Science and Technology Facilities Council, RAL Space, Rutherford Appleton Laboratory, Harwell Campus, Didcot, OX11 0QX, UK
[3]Departamento de Física, Universidad de Extremadura, E-06800 Mérida, Spain
[4]Research Institute for Sustainable Humanosphere, Kyoto University, Uji, 6100011, Japan
[5]Unit of Synergetic Studies for Space, Kyoto University, Kyoto, 6068306, Japan

*Correspondence to*: Hisashi Hayakawa (hayakawa@kwasan.kyoto-u.ac.jp)

**Abstract.** While low latitude auroral displays are normally considered to be a manifestation of magnetic storms of considerable size, Silverman (2003, *JGR*, 108, A4) reported numerous "sporadic auroras" which appear locally at relatively low magnetic latitudes during times of just moderate magnetic activity. Here, a case study is presented of an aurora near the geomagnetic equator based on a report from the Philippine Islands on 27 October 1856. An analysis of this report shows it to be consistent with the known cases of sporadic aurorae except for its considerably low magnetic latitude. The record also suggests that extremely low-latitude aurora is not always accompanied with large magnetic storms. The description of its brief appearance leads to a possible physical explanation based on an ephemeral magnetospheric disturbance provoking this sporadic aurora.

**1 Introduction**

It is known that a low-latitude aurora is a manifestation of a magnetic storm caused by solar eruptions (e.g., Gonzalez et al., 1998; Shiokawa et al., 2005; Willis et al., 2006; Odenwald, 2015). Since the beginning of modern magnetic observations in the mid-19th century, magnetic records have been compared with auroral displays (e.g., Allen et al., 1989; Silverman, 1995, 2006, 2008; Silverman & Cliver, 2001; Shiokawa et al., 1998, 2005; Vaquero et al., 2008). In August and September 1859, solar eruptions from large sunspots caused an intense magnetic storm reaching values as extreme as 1600 nT in the horizontal geomagnetic field at Colaba (Tsurutani et al., 2003; Nevanlinna, 2006; Ribeiro et al., 2011), with major auroral displays seen worldwide down to magnetic latitudes (hereafter, MLATs) as low as ~20° (Kimball, 1960; Cliver & Svalgaard, 2004; Green & Boardsen, 2006; Farrona et al., 2011; Cliver & Dietrich, 2013; Hayakawa et al., 2016b; Lakhina & Tsurutani, 2016).

However, it is reported that auroral displays at low MLATs also occur during low or moderate geomagnetic disturbances. Silverman (2003) examined these auroral displays at relatively low MLAT during low or moderate geomagnetic disturbances in the *Climatological Data of the United States* during 1880 to 1940, identifying 54 cases in the United States, and attesting to the reality of "sporadic aurorae", using the terminology of Botley (1963) who defined this phenomenon as a "single ray in a sky otherwise seemingly clear of auroral light, or isolated patches well to the equatorial side of a great display" citing Abbe (1895).

Willis et al. (2007) and Vaquero et al. (2007, 2011) surveyed this kind of localized low-latitude auroral display in China, Spain, and Mexico to identify reports during low or moderate geomagnetic activity. Silverman (2003) and Willis et al. (2007) drew attention to the question of the mechanism behind them, as to how the localized auroral display can be seen at a low latitude without there being any intense magnetic storms.

In this short contribution, we aim to describe a case of a "sporadic aurora" in the Philippine Islands, close to the geomagnetic equator. It should be noted that aurorae near the geomagnetic equator have yet to be studied, and knowledge of them will be an important key to scientific understanding of "sporadic aurorae".

## 2 Material and Method

Antonio Llanos (1806-1881), a Spanish priest with interest in botany and meteorology (Vaquero et al., 2005), reports a curious account of an "Observation of an aurora borealis in Manila (*Observación de una aurora boreal en Manila*)" (Llanos, 1857). As is explicit in the title, Llanos considered this phenomenon an "aurora borealis" while being aware that the appearance of an aurora at such low latitudes is extremely rare. He associates the appearance of this aurora to exceptional (and unknown) circumstances of the atmosphere, and therefore writes up this report so that physicists working on the origin of the phenomenon shall have evidence of this unusual observation.

Based on this historical report by Llanos, we shall consider the nature of this phenomenon, compute the contemporary MLAT of the observation site, and compare the record with contemporary geomagnetic activity. Magnetic observations started in the 1840s, and the *ak* index is available from 1844 onwards, while the *aa* index is available from 1868 (Nevanlinna, 2004; Willis et al., 2007). We examine the values of the *ak* index (Nevanlinna & Kataja, 1993; Nevanlinna, 2004) around the date of observation provided by Llanos.

# 3 The Aurora Borealis on 27 October 1856

Antonio Llanos reported the auroral display to a Spanish journal entitled *Revista de los Progresos de las Ciencias Exactas, Físicas y Naturales* (see Fig. 1). We shall summarize his report and review his observation. First, we shall extract Llanos's description of the observational report:

**Figure 1**. The original report in Spanish by Llanos (1857).

"At this moment [at 9 o'clock at night], observing the cloudscape of the atmosphere, I noticed that, on the NW side, with a short difference there was a faint but weak white light on that horizon, which at first I supposed was produced by some cause, such as from a fire. In that part, there is a range of mountains that form the provinces of Balanga and Zambales. The illuminated space would only rise about 4° above the horizon, and the segment width would be about 25°. It seemed to be on the skirt or side of these mountains opposite the NW, and as if it were stopped there, prevented its passage by the said mountain ranges. At its base, the light was noticed to be more clear and perceptible, and some more resplendent points could be seen in its mass, noting also some movement of vertical undulation which it manifested, sometimes stronger and sometimes weaker, until finally it disappeared, leaving total darkness. When I began to notice it, I found it in the said state, and the time of duration in my view would be some 5 minutes. That illumination had scarcely disappeared, when on the opposite side of the first quadrant, that is, in the NE, the same phenomenon was repeated with the same circumstances as the previous one, although with a greater extension, there being also another mountain range called Gapang, which runs in the same direction from N to S, finding myself in the basin that these two ranges comprise; but on this occasion it lasted longer, or double the first, and it was 10 minutes, with the wind firmly on the same side or a little more to the E, and with quite a lot of rain."

**4 The Observational Site and its Magnetic Latitude**

Antonio Llanos explicitly writes his observational site as being Manila, and its geographical latitude as at "latitude 15°N, a little more or less". We estimate his observational site as the city centre of Manila (14°35' N, 120°58' E). We computed the contemporary MLAT for this place in 1856, based on the dipole component of the GUFM1 geomagnetic field model (Jackson et al., 2000). We obtained the value of 3.3° MLAT. This value is within 0.05° of difference from that in 1900 as computed by the IGRF model (Thébault et al., 2015). Therefore, one can fairly consider this observation to have been made near the geomagnetic equator.

It is not common for auroral displays to be seen anywhere near the geomagnetic equator. In some extreme magnetic storms, it is known that auroral displays were visible down to some 18° to 30° MLAT, such as those in the major storms of 1989, 1921, 1909, 1872, 1870, 1859, 1770, and 1730 (Kimball, 1960; Allen et al., 1989; Silverman, 1995, 2006, 2008; Silverman & Cliver, 2001; Vaquero et al., 2008; Hayakawa et al., 2017, 2018a, 2018b; Ebihara et al., 2017; Willis et al., 1996), as partially reviewed by Cliver & Svalgaard (2004) and Cliver & Dietrich (2013). However, this value (3.3° MLAT) is evidently closer to the geomagnetic equator, and is much lower than in the other events.

**5 Nature of this Phenomenon**

It is worth consideration as to whether this record of an "aurora borealis" can be related to other phenomena. Antonio Llanos suspected this phenomenon at first to be "as from a fire", and ended by describing it as a "meteor that is so rare at low northern latitudes" following his conclusion that it was indeed an "*aurora boreal*". Nonetheless, it is possible to find atmospheric optics or comet tails have been misinterpreted as auroral displays (e.g., Hayakawa et al., 2015, 2016a; Kawamura et al., 2016; Carrasco et al., 2017; Usoskin et al., 2017).

Its colour was described "white" and less like low latitude auroras. However, due to the Purkinje Effect, the human eyes frequently perceive weak lights as apparently whitish, as they are insensitive to color with weak brightness (Purkinje, 1825, p.109; Minnaert, 1993, p.133). Moreover, it was described "a faint but weak white light on that horizon" and hence its brightness is considered rather faint and weak. Therefore, it is likely that this phenomenon is perceived *apparently* whitish due to the Purkinje Effect.

Atmospheric optics is dependent on the Moon for its light source (e.g., Minnaert, 1993). We computed the lunar phase on 1856 October 27, and obtained a value of 0.96 based on the method described by Kawamura et al. (2016) developed from Meeus (1988). This means that it was almost a new moon, and one can probably exclude the possibility that the light was associated with atmospheric optics from moonlight at night. Fogbows cannot explain this phenomenon either as they have a width of 25° or greater, while much smaller than normal rainbows, and they appear "nearly always … when the dazzling beam of a car's headlights behind you penetrates the mist in front of you" (Minnaert, 1993, pp.201-202). Llanos did not describe any such "dazzling beam behind" him.

Likewise, its description of "width of 25° or greater" and duration for "some 5 minutes" or "10 minutes" show us an upward discharge from the top of thundercloud is also unlikely (e.g. Pasko et al., 2002), considering this glow was seen beyond the mountain ranges of Balanga and Zambales, about 60 km and 140 km away from Manila respectively.

We also considered the possibility of a meteor shower. Within the October meteor showers listed in the catalogue of Kronk (2014, pp.227-255), the Orionids are one of the candidates. However, Llanos reported "At its base, the light was noticed to be more clear and perceptible", and it is unlikely that a meteor shower will decrease in brightness near the horizon. Moreover, the duration of 5 or 10 minutes is too short for a meteor shower. Likewise, it is also difficult to consider that this phenomenon might have been a comet tail as it lasted only 5 minutes in the NW and 10 minutes in the NE. Neither does Kronk (2003, pp.245-246) report any comets in late 1856.

Mountain fire is also unlikely. While Llanos first suspected a fire in the mountains to be the cause, he had not got any reports of fire in the northern mountains of Manila at least until his publication. This phenomenon had a width of 25° or greater and it would thus have to have been a large fire, which would have soon been reported to Manila if it were a fire in the mountains. Auroral displays are frequently mistaken for conflagrations when they are bright enough. In the Carrington event, a considerable number of observers in East Asia and North America misinterpreted the auroral displays as being conflagrations (Green et al., 2006; Hayakawa et al., 2016b). Similar reports are found during other large magnetic storms with bright auroral displays (Silverman, 2008; Odenwald, 2007; Vaquero et al., 2008; Ebihara et al., 2017; Hayakawa et al., 2017).

It seems therefore that one has no strong reason to reject this as being one instance of "sporadic aurorae" which appear locally at relatively low MLAT, as reported in Silverman (2003). This case had a horizontal appearance, of ~25° in width and 4° in elevation. We would also note that it appeared in the NW direction for 5 minutes, and then in the NE direction for 10 minutes. Its base was brighter than the upper part, with "vertical undulation". These features also suggest its being interpreted as a kind of auroral display. Assuming that the altitude of the upper part of the aurora was 400 km, we estimated that the aurora would have appeared at 19.5° MLAT (23.9° invariant latitude, ILAT, in the magnetic coordinates used to specify a magnetic field line in the space physics community). ILAT $\Lambda$ is constant along a field line, and is given by

$$\Lambda = \cos^{-1}\left(\sqrt{1/L}\right),$$

where $L$ is the distance in units of the Earth's radius between the centre of the Earth and the point where the magnetic field line crosses the equatorial plane (McIlwain, 1966). In contrast, MLAT $\lambda$ varies along a field line, and is given by

$$\lambda = \cos^{-1}\left(\sqrt{R/L}\right),$$

where $R$ is the distance between the centre of the Earth and the specific point. At the surface of the Earth, $\Lambda$ is equal to $\lambda$.

**6 Contemporary Solar and Geomagnetic Activities**

It is intriguing where this event is situated relative to solar and geomagnetic activities. It is known that the frequency of occurrence of magnetic storms is in relatively good agreement with the sunspot number (e.g., Willis et al., 2006; Vázquez et al., 2006), and recent statistical studies reveal that even the quieter Sun can on occasion also cause superstorms (e.g., Kilpua et al., 2015).

In terms of long-term solar activity, this event was mostly situated near the solar minimum in 1856 (e.g., Clette et al., 2014; Vaquero et al., 2016). The solar surface in October 1856 showed only a few sunspots (Carrington, 1863; Vaquero et al., 2016). Figure 2 shows the daily *ak* value observed at Helsinki according to Nevanlinna (2004), indicating that the geomagnetic activity was also very low. Figure 3 shows the H-component of the geomagnetic field disturbances (ΔH) with a 1-hour resolution. In the second half of the 19th century, a typical precision of a magnetometer is around 1' (e.g. Batlló, 2005) and may have caused apparently large pseudo-random variations than those in the modern time. On 27 October 1856, ΔH at the Helsinki observatory (geographic latitude 60.2° and geographic longitude 25.0°) exhibits a negative excursion, peaking at 15 UT, with an amplitude of ~140 nT as shown in Figure 3a. The sporadic aurora occurred around 21:00-21:15 LT (12:56-13:11 UT) at Manila, which is roughly corresponding to the descending phase of this negative excursion at Helsinki, by considering that the differences of time zones between between Manila (N14°35', E120°58'), Helsinki observatory (N60°10′, E24°57′), and Greenwich are roughly 7.07h and 8.06h on the basis of local mean time (e.g. Nevanlinna, 2006, 2008). If this negative excursion is caused by the ring current, the secular variation is negligible, and the magnetic disturbance is independent of the magnetic local time, then the Dst would be calculated approximately as Dst = ΔH/cos λ, where ΔH is the magnetic disturbance (Sugiura, 1964). Substituting ΔH of ~140 nT and λ of 58.2° (Helsinki observatory), we estimated Dst to be ~−266 nT. The recovery of the negative excursion takes place for only 1 hour, which is too short to attribute to the decay of the storm-time ring current (Ebihara & Ejiri, 2003). The ionospheric current could also contribute to the variation of ΔH. Fig. 3b shows ΔH at the Lovö observatory (geographic latitude of 59.3° and geographic longitude 17.8°) in the March 1989 storm. The Lovö observatory is close to Helsinki. To date, the March 1989 storm is the largest since 1957 in terms of the minimum Dst values (-589 nT.) The amplitude of ΔH exceeds 1000 nT, which is probably associated with the ionospheric current (in addition to other current systems such as the ring current), and is much larger than observed in Helsinki on 27 October 1856. Although the cause of the magnetic disturbance is uncertain, it can be said that the magnetic disturbance on 27 October 1856 was most likely low, at least at Helsinki, in comparison with the large storm in March 1989. Figure 3c shows ΔH at the Lovö observatory on 17-21 January 2002. The variation of ΔH on 19 January 2002 resembles that observed on 27 October 1856 in terms of the negative excursion and subsequent variation. The negative excursion starts at ~12 UT, and peaked at ~16 UT on 19 January 2002. According to the OMNI solar wind data (King and Papitashvili, 2005), the negative excursion is associated with a southward turning of the interplanetary magnetic field (IMF) and a rapid increase in the solar wind dynamic pressure (data not shown). The sudden increase in the solar wind dynamic pressure resulted in the sudden increase in ΔH, which is visible in the one-minute resolution data at Lovö (dotted line in Figure 3c). The southward IMF continued until ~15 UT, which could result in the intensification of the ring current, and the negative variation of ΔH. ΔH is highly fluctuating throughout this period, which is caused by fluctuations of the solar wind and IMF. The solar wind speed and density increased gradually, starting at ~05 UT on 19 January 2002, and the strength of IMF peaked at ~9 UT on 19 January 2002. These characteristics may correspond to a corotating interaction region (CIR) (Denton et al., 2006). The Dst index did not reach −30 nT on 19-20 January 2002. The amplitude of the negative excursion (~140nT) observed in 1856 is roughly 5 times larger than that observed in 2002 (~-30nT). This might indicate that the IMF Bz and/or solar wind velocity in 1856 was larger than those in 2002.

Therefore, we cannot find evidence of any strong geomagnetic disturbance on 27 October 1856 as in intense magnetic storms such as the superstorms in 1859 that brought auroral display down to low MLAT (Kimball, 1960; Tsurutani et al., 2003; Cliver & Dietrich, 2013). One possible scenario is that a short-lasting magnetospheric disturbance occurred to cause the sporadic aurora. The disturbance is probably associated with a rapid enhancement of the magnetospheric electric field which transports magnetospheric electrons deeply earthwards (inwards). After being rapidly transported, the electrons were probably scattered by some processes on the field line at the *L*-value of 1.20 (23.9° ILAT). The scattered electrons could then have precipitated into the upper atmosphere, exciting oxygen atoms so as to cause the aurora. The disturbance should have been strong, at least at the *L*-value of 1.20, but the duration should have been short (within at most 15 minutes). If the duration of a strong disturbance (convection) is relatively long, hot ions also move inwards so as to intensify the plasma pressure (the ring current) that principally disturbs the geomagnetic field characterized by a negative excursion of the H-component of the magnetic field (Ebihara and Ejiri, 2003). The observation shows that the ring current was not strongly developed during this period. One of the possible causes for the short-lasting, large-amplitude disturbance is the interplanetary shock that reached the Earth. The compressional magnetospheric wave that was excited at the dayside magnetopause could propagate towards the Earth in the direction perpendicular to the magnetic field (e.g., Wilson & Sugiura, 1961). Shock-associated disturbances are observed in the magnetosphere at all magnetic local times at *L*-value as low as ~1.2 (Shinbori et al., 2003, 2004). The transient compression of the magnetic field in the magnetosphere could result in the excitation of electromagnetic ion cyclotron (EMIC) waves (e.g., Immel et al., 2005) and chorus waves (e.g., Fu et al., 2012; Zhou et al., 2015). Interacting with the EMIC or chorus waves, the magnetospheric particles undergo pitch angle scattering, resulting in their precipitation into the upper atmosphere. According to observations, the wave intensifications and shock-associated aurorae occur primarily on the dayside (e.g., Anderson & Hamilton, 1993; Zhang et al., 2004; Zhang et al., 2008; Zhou et al., 2015). This seems to be inconsistent with the present aurora observation which was made at 9 o'clock at night, local time. If the normal angle of the shock slants a lot, the impact of the interplanetary shock could be large enough in the late evening region (e.g., Selvakumaran et al., 2017) to excite EMIC and/or chorus waves at probably 9 o'clock at night, local time.

Usually, the magnetic disturbance associated with an interplanetary shock lasts for just a few minutes ~ a few tens of minutes depending on solar wind dynamic pressure (Araki et al., 2004) and orientation angle of the shock front (Takeuchi et al., 2002). This short duration may explain why no significant disturbance was recorded in the daily *ak* index as shown in Fig. 2, and in the hourly geomagnetic field data at Helsinki (N60°10′, E24°57′) as shown in Fig. 3a. Since shock-associated magnetic disturbance is a global phenomenon (e.g., Nishida and Jacobs, 1962; Araki, 1994), the disturbance would have been detectable at Helsinki if the temporal resolution was high enough as shown in Figure 3c. Due to its short duration, other observers may have missed it, instead seeing the clear sky at around "9 o'clock at night", Manila local time. This may explain why we have no auroral report on that same night at around 23.9° ILAT, for example, from observers in East Asia (Willis et al., 2007).

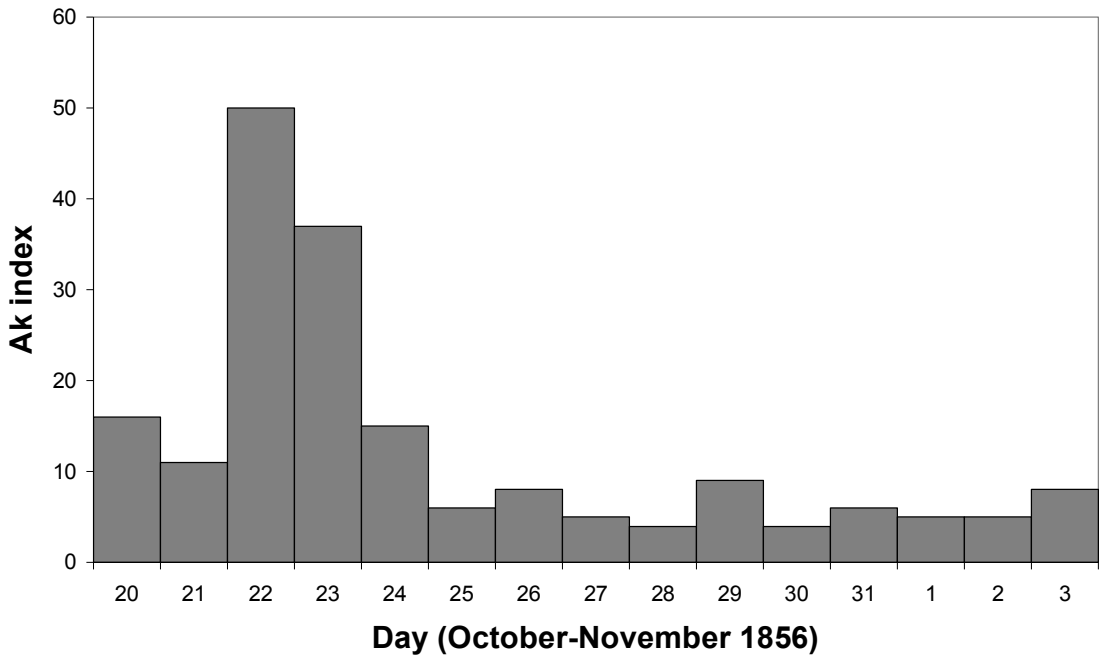

**Figure 2.** Daily *ak* index (Nevanlinna, 1997) during the period 20 October - 3 November 1856.

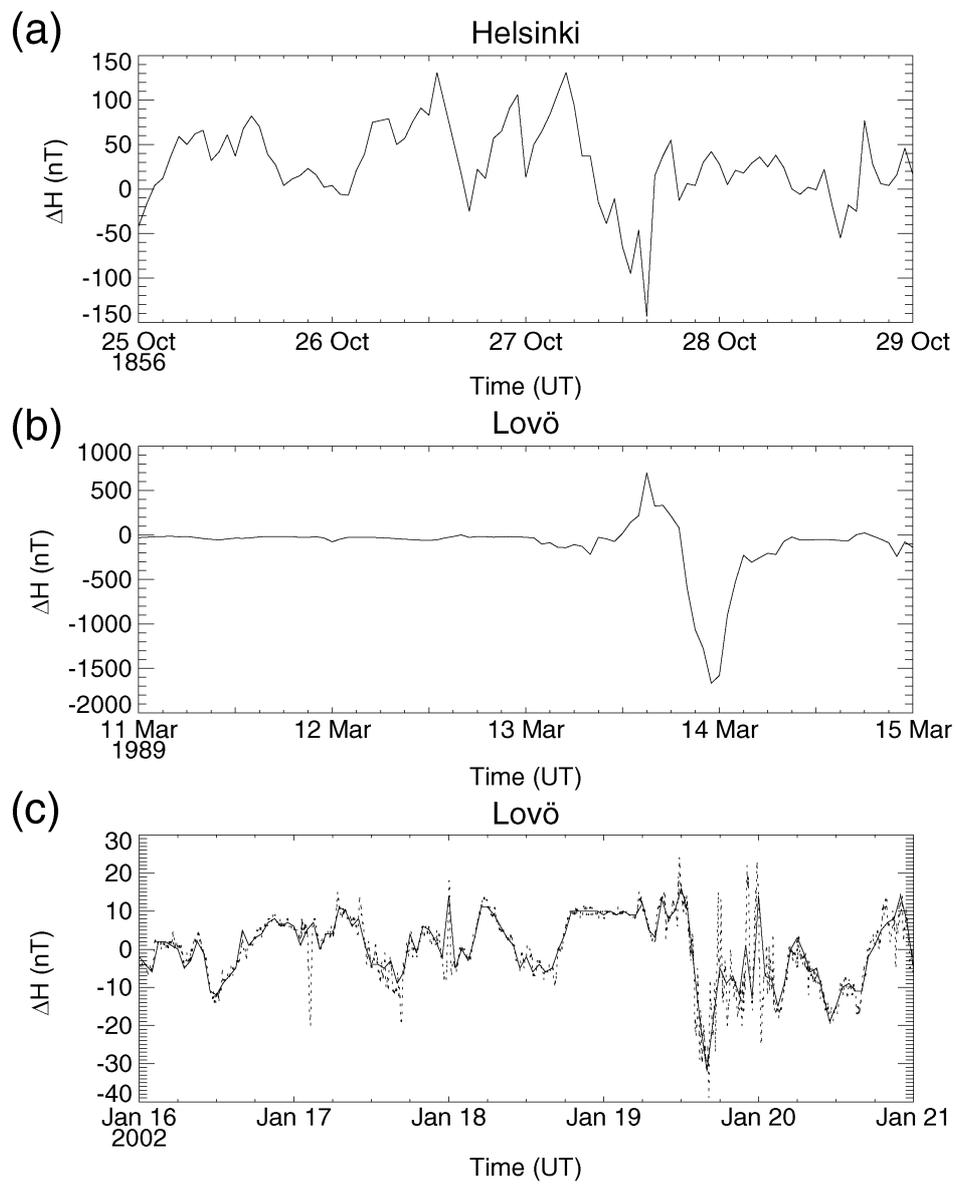

**Figure 3**. From top to bottom, the H-component of the geomagnetic field disturbance at Helsinki in 1859, Lovö in 1989, and Lovö in 2002. The dotted line indicates the one-minute data.

# 7 Conclusion

In this short contribution, we have examined the record of an "aurora borealis" at Manila on 27 October 1856. According to our analysis of the text, we consider this record to indeed be likely one of an auroral display as was considered by the observer himself, Antonio Llanos. Reconstruction of contemporary MLAT showed that Manila was situated at 3.3° MLAT, close to the geomagnetic equator. However, we could find no large sunspots or large geomagnetic storms associated with this auroral report. We did not find any contemporary auroral display reports in Willis et al. (2007). This means that this auroral display was local at a low MLAT, and should be categorized as an instance of "sporadic aurorae". On the analogy to the magnetic variation observed at Lovö in 2002, the sporadic aurora may be associated with a shock embedded in an interface of corotating interaction region (CIR). The shock may result in transmission of an electromagnetic pulse propagating in the magnetosphere. In the course of the propagation, magnetosphereic electrons could precipitate into the ionosphere, brightening the sporadic aurora. Further studies are needed to confirm this scenario in the future. As far as we know, this example is the first evidence for a sporadic aurora in South East Asia and near the geomagnetic equator. Together with known records of sporadic aurorae from the United States (Silverman, 2003), East Asia (Willis et al., 2007), Spain (Vaquero et al., 2007), and Mexico (Vaquero et al., 2011), this record should provide a further resource with which to consider the physical nature of this phenomenon. Although this is rather an isolated phenomenon, further researches for this phenomenon may merit studies of long-term variations of geomagnetic activity and terrestrial magnetic field as well.

**Acknowledgement**

The authors are indebted to Heikki Nevanlinna and Ari Viljanen who have provided the daily *ak* index values and World Data Center for Geomagnetism, Kyoto for providing the magnetic observation data. This research was also partially supported by the Economy and Infrastructure Board of the Junta of Extremadura through project IB16127 and grant GR15137 (co-financed by the European Regional Development Fund), the Ministerio de Economía y Competitividad of the Spanish Government (AYA2014-57556-P and CGL2017-87917-P), grant-in-aid from the Ministry of Education, Culture, Sports, Science and Technology of Japan, grant numbers JP15H05816 (PI: S. Yoden), JP15H03732 (PI: Y. Ebihara), JP16H03955 (PI: K. Shibata), JP 18H01254 (PI: H. Isobe), and JP15H05815 (PI: Y. Miyoshi), and a grant-in-aid for JSPS Research Fellow JP17J06954 (PI: H. Hayakawa), and the Exploratory and Mission Research Projects of the Research Institute for Sustainable Humanosphere (PI: H. Isobe). The OMNI data were obtained from the GSFC/SPDF OMNI Web interface at https://omniweb.gsfc.nasa.gov. The authors gratefully thank Dr. Sam M. Silverman for attracting our attention to the sporadic aurora and Dr. Tiera Laitinen and another anonymous referee for their helpful and constructive comments on our paper.

Zhou, C., Li, W., Thorne, R. M., et al.: Excitation of dayside chorus waves due to magnetic field line compression in response to interplanetary shocks, *J. Geophys. Res. Space Physics*, 120, 8327–8338, doi:10.1002/2015JA021530, 2015.